\documentclass[10pt, conference]{IEEEtran}
\IEEEoverridecommandlockouts
\usepackage{cite}
\usepackage{booktabs}
\usepackage{subcaption}
\usepackage{url}
\usepackage{amsmath,amssymb,amsfonts}
\usepackage{algorithmic}
\usepackage{graphicx}
\usepackage{textcomp}
\usepackage{xcolor}
\usepackage{verbatim}
\usepackage{multirow}
\usepackage{array}
\usepackage{quantikz}
\usepackage{adjustbox}
\usepackage{caption}
\usepackage{tikz}
\usepackage{hyperref}

\def\BibTeX{{\rm B\kern-.05em{\sc i\kern-.025em b}\kern-.08em
    T\kern-.1667em\lower.7ex\hbox{E}\kern-.125emX}}
\begin{document}

\title{Influence of Data Dimensionality Reduction Methods on the Effectiveness of Quantum Machine Learning Models \\ 
}

\author{\IEEEauthorblockN{Aakash Ravindra Shinde}
\IEEEauthorblockA{\textit{Department of Computer Science} \\
\textit{University of Helsinki}\\
Helsinki, Finland \\
shinde.aakashravindra@helsinki.fi}
\and
\IEEEauthorblockN{ Jukka K. Nurminen}
\IEEEauthorblockA{\textit{Department of Computer Science} \\
\textit{University of Helsinki}\\
Helsinki, Finland \\
jukka.k.nurminen@helsinki.fi}

}

\maketitle

\begin{abstract}
Data dimensionality reduction techniques are often utilized in the implementation of Quantum Machine Learning models to address two significant issues: the constraints of NISQ quantum devices, which are characterized by noise and a limited number of qubits, and the challenge of simulating a large number of qubits on classical devices. It also raises concerns over the scalability of these approaches, as dimensionality reduction methods are slow to adapt to large datasets. 
In this article, we analyze how data reduction methods affect different QML models. We conduct this experiment over several generated datasets, quantum machine algorithms, quantum data encoding methods, and data reduction methods. All these models were evaluated on the performance metrics like accuracy, precision, recall, and F1 score. Our findings have led us to conclude that the usage of data dimensionality reduction methods results in skewed performance metric values, which results in wrongly estimating the actual performance of quantum machine learning models. There are several factors, along with data dimensionality reduction methods, that worsen this problem, such as characteristics of the datasets, classical to quantum information embedding methods, percentage of feature reduction, classical components associated with quantum models, and structure of quantum machine learning models. 
We consistently observed the difference in the accuracy range of 14\% to 48\% amongst these models, using data reduction and not using it. Apart from this, our observations have shown that some data reduction methods tend to perform better for some specific data embedding methodologies and ansatz constructions.
\end{abstract}


\section{Introduction}
In recent decades, there has been a significant push towards research and development of Quantum Machine Learning algorithms and models. Quantum Machine Learning has also been heralded as one of the prominent use cases for Quantum Computing devices. Several studies have shown the ability of QML models to solve difficult machine-learning problems and sometimes outperform the classical approach. Mostly, these proofs are either theoretical or simulated on classical devices. This is because the current quantum computational devices lack the required number of qubits, have questionable error correction ability, and tend to have noisy qubits. Both the theoretical approach and classical simulations have their qualms. 

If the QML model is supported only by theoretical proof, it often relies on assumptions. These include the existence of optimizers that perform ideally under the model's setup and access to resources like quantum RAM, which are assumed to provide speedups and support the theoretical claims \cite{Aaronson:2015scy}.
While such theories are vital for the progress of QML as a viable application for quantum computing, their ability to be implemented should be questioned. On the other hand, classical simulation of quantum systems also has its set of problems. The biggest one is the inability of classical computational devices to simulate larger quantum systems. This is a critical issue for QML as this makes it impossible to simulate a large number of qubits and represent datasets with a higher number of feature values. To circumvent this issue, such classical simulations are performed using small toy datasets with a low feature number, or they use data dimensionality reduction techniques for feature reduction on larger datasets; in a few cases, both approaches are used.  

So the question arises: why is the use of data dimensionality reduction techniques a concern for Quantum Machine Learning implementations? Firstly, it creates a scalability issue for quantum machine learning algorithms. As the size of the data increases, i.e., the number of data points increases, the computational cost and the memory cost for implementing these data dimensionality reduction methods increase accordingly. These dimensionality reduction methods are not only expensive from a time complexity perspective but also from a space complexity perspective \cite{van2009dimensionality}. This, in turn, results in slowing down the overall implementation of the Quantum Machine Learning algorithm. 

The potential of Quantum Machine Learning (QML) as a viable application in quantum computing is often attributed to its theoretical advantages in processing speed compared to classical methodologies. However, this perceived speedup may be rendered insignificant if the techniques employed for dimensionality reduction contribute to an increase in overall complexity. Secondly, one cannot estimate the true performance of the proposed Quantum Machine Learning algorithms and provide proof that they will perform similarly when scaled up. 

In such a case, one might think of implementing proposed QML algorithms on non-reduced toy datasets. The toy datasets come with their own set of problems, as they miss several factors one might observe in real-world datasets and lack inherent complexity. This might prove the QML implementations to be underperforming in such real-world dataset scenarios. Hence, these models need to be verified on datasets with more variance, noise, and other factors. Bowles et al. \cite{bowles2024betterclassicalsubtleart} provide a set of generated datasets to verify and validate current QML models. This research mentions the idea of consistent use of the data dimensionality reduction approach for more than half of the randomly selected papers. Yet, they tend to overlook such implementations. We analyzed the most cited (''Quantum Machine Learning'') papers list from the 'Web of Science' website and the papers with more than 30 citations in the ' Quantum Machine Intelligence' Journal, and saw around 23 papers with similar trends. 

From this list of papers, we saw that some of them only had a theoretical proof without any experimental results, some were either simulated classically, and very few on a noisy quantum device. Many who did do simulations tend to restrict their results to datasets with a really small number of features. The rest of them catered to use either datasets that were already data dimension reduced, like an $8\times8$ MNIST, or used data dimension reduction techniques to fit the values to the limited number of qubits \cite{Shinde_QuantumMachineLearningReviewPaperList_2025}.

To address these concerns, we have designed this experiment with varying kinds of quantum ansatz to have at least some resemblance to most of the proposed quantum machine learning techniques. We further evaluate these models based on different generated datasets that have the ability to add noise, redundancy, and variance. We also consider and evaluate the effects of data reduction on the percentage of features reduced. There are a few constraints kept constant for the comparison to be fair. We decided to create the largest models based on the maximum number of feasible simulable qubits and the depth in consideration. For this experiment, we addressed the following research questions:

\begin{itemize}
  \item How are different Quantum Machine Learning Models affected by the use of data dimensionality reduction methods?
  \item Does the use of different Embedding methods change the effects of data dimensionality reduction?
  \item Does the different percentage of knowledge distillation result in a change in overall performance?
\end{itemize}

This manuscript follows the following structure. In Section \ref{DataRedu} and \ref{QML models}, we discuss the basic ideas surrounding Data Dimension reduction and Quantum Machine Learning models, respectively. We also discuss the algorithms and their implementations that we used for these experiments. Section \ref{Dataset} describes the construction and the reasoning for choosing the datasets to experiment over and the tunable parameters that make them unique, trying to cover as much ground as we can. Section \ref{Experimental Setup} provides a deeper understanding of the Experimental setup and how all these factors fit in together to provide a fair comparison. Finally, Section \ref{Results} provides are findings and experimental observations. These results are further discussed, interpreted, and validated in Section \ref{Discuss}.

\section{Data dimensionality Reduction}
\label{DataRedu}
Data dimensionality reduction algorithms are fundamentally responsible for transforming high-dimensional data into a meaningful representation of reduced dimensionality, especially correlated to the intrinsic data dimensionality. Data reduction methods can effectively gain significant accuracy in classical ML models as presented in \cite{ReductionTechniquesonBigData}\cite{DimensionalityReductionTec}. There is always the concern that the data reduction methods can result in information loss if all parameters are equally weighed or if some parameters hold niche information \cite{van2009dimensionality}. 

One method for classifying data reduction methods is based on their transformation, i.e., linear and non-linear transformation methods. Linear data dimensionality reduction methods employ techniques to reduce the dimensions to a lower-dimensional subspace using a linear combination of the original data values. They work on the assumption that the data lies on or near a linear subspace of a high-dimensional subspace. However, linear techniques lack the ability to work effectively on non-linear data. 

\begin{figure}
  \centering
  \includegraphics[width=0.40\textwidth]{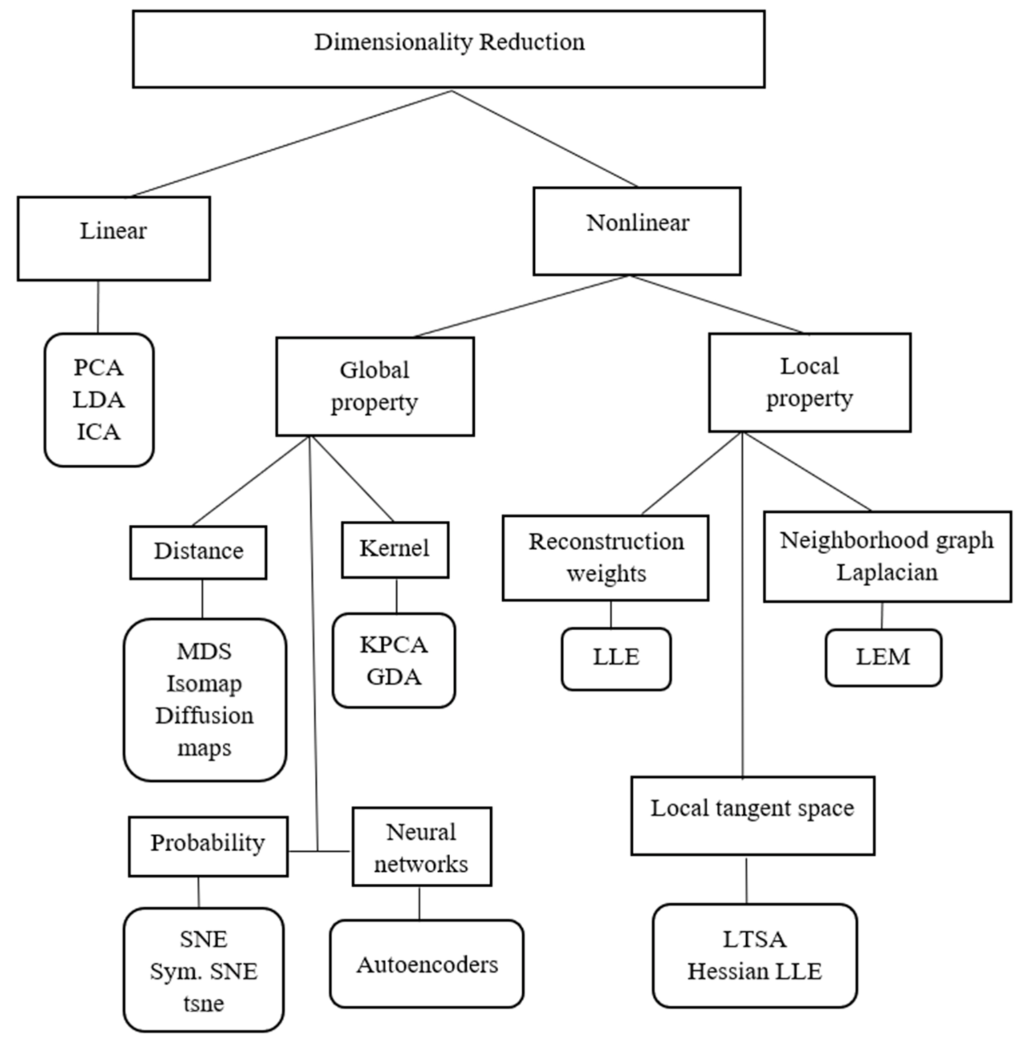}
  \caption{Taxonomy of Data Reduction Methods by \cite{C2023PerformanceEO}}
  \label{fig:dim_tree}
\end{figure}

Unlike linear techniques, non-linear dimensionality reduction methods aim to reduce the dimensionality of complex, non-linear data by mapping it from a high-dimensional space to a lower-dimensional space. Non-linear dimensionality reduction methods can be broadly characterized as selection-based and projection-based techniques. Non-linear techniques can be classified further into those that preserve the global properties of the original data in a lower dimension, those that preserve local or niche properties, and those that perform global alignment of a mixture of linear models \cite{van2009dimensionality}. Figure \ref{fig:dim_tree} represents a tree structure for the general classification of data dimensional reduction techniques presented in \cite{C2023PerformanceEO}.

 Van Der Maaten et al. \cite{van2009dimensionality} provides great insights into a few general properties of data dimensionality reduction methods and their pros and cons associated with the mentioned properties. The first property discussed in the article is about the parametric nature of the mapping from a higher dimension to a lower-dimensional space. Most of the dimensionality reduction methods are non-parametric, which means they do not have a defined mapping procedure. A key drawback of non-parametric methods is that they do not easily generalize to new or unseen test data without reapplying the dimensionality reduction process. Additionally, it is difficult to assess how much of the original high-dimensional information is preserved in the low-dimensional representation since reconstructing the original data and measuring the reconstruction error is not straightforward \cite{van2009dimensionality}. 

The second property discussed is the availability of free tunable parameters for optimization, which is generally a characteristic of non-linear methods. Free tunable parameters come with both benefits and drawbacks. On the positive side, they offer greater flexibility to the dimensionality reduction technique. However, the downside is that these parameters must be carefully tuned to achieve optimal performance. Finally last two properties explored were regarding the Space and Time complexity of the reduction methods, which is relative to each of the methods.

We considered working with four specific data reduction methods, which are Principal Component Analysis (PCA), Truncated Support Vector Decomposition, Autoencoder, and t-distributed stochastic neighbor embedding(t-SNE). PCA and Truncated SVD represent the linear dimensionality reduction methods, whereas Autoencoder and t-SNE represent the non-linear dimensionality reduction methods. Analysis of these four methods based on the properties is provided in Table \ref{tab: property_anal}.

\begin{table}[ht!]
    \centering
    \begin{tabular}{|c|c|c|c|c|}
    \hline
     \textbf{{\parbox{1cm}{\centering Methods}}} & \textbf{{\parbox{1.3cm}{\centering Parametric}}} & \textbf{{\parbox{1.2cm}{\centering Tunable Parameters}}} & \textbf{{\parbox{1.2cm}{\centering Time Complexity}}} & \textbf{{\parbox{1.2cm}{\centering Space Complexity}}} \\
       \hline
        PCA & no & none & $O(n^2)$ & $O(m^2)$ \\
        Autoencoder & yes & none & $O(inw)$ & $O(w)$ \\
        t-SNE & no & yes & $O(n^2)$ & $O(m^2)$ \\
        Truncated-SVD & no & none & $O(n^3)$ & $O(m^2)$ \\
    \hline
    \end{tabular}
    \vspace{3pt}
    \caption{Property analysis of Reduction Methods. Here, $n$ defines the number of data points, $m$ the dimensionality, $i$ the number of iterations, and $w$ the number of weights in the neural network \cite{vanDerMaaten2008,van2009dimensionality,li2019tutorialcomplexityanalysissingular}.}
    \label{tab: property_anal}
\end{table}

\subsection{Principal Component Analysis (PCA)}
PCA is one of the most popular linear (unsupervised) dimensionality reduction techniques and has been extensively used in several manuscripts for training QML models \cite{Shinde_QuantumMachineLearningReviewPaperList_2025}. PCA works on the principle of creating lower-dimensional data that describes as much variance as possible. Mathematically, PCA attempts to find a linear map $M$ that maximizes $M^T cov(X) M$, where $cov(X)$ is a covariance matrix of data $X$. It can further be shown that this linear map consists of $k$ principal eigenvalues, i.e., principal components of the covariance matrix of the zero-mean data. Hence, PCA solves the eigenproblem $cov(X)M = \lambda M$ \cite{van2009dimensionality}.

\subsection{Truncated Singular Value Decomposition (Truncated SVD)}
Truncated SVD is a variant of SVD that computes the $d$ largest singular values, where $d$ is the required reduced dimension. Truncated SVD shares a lot of similarities with PCA but differs in the idea of Matrix $X$ not being centered. Mathematically, truncated SVD, when applied to a training sample $X$, produces a low-rank approximation $X$ such that $X \approx X_d = U_d \Sigma_d V_d^T$. Here, $U$ is a $m \times m $ complex unitary matrix, $\Sigma$ is an $m \times n$ rectangular diagonal matrix with non-negative real numbers on the diagonal, $V$ is an $ n \times n$ complex unitary matrix, and $V^T$ is the conjugate transpose of $V$. After this operation, $U_d\Sigma_d$ is transformed training set with $d$ features \cite{scikitTruncatedSvd}. 

\subsection{Multilayer autoencoder (Autoencoder)}
Multilayer autoencoders are odd-layered feed-forward neural networks with input layer dimensions equivalent to the original data dimensions and the output layer equal to the desired dimensions with several hidden middle layers. These networks are trained with the motif of reducing the mean squared error between the input and output values \cite{van2009dimensionality}. This process is conducted repetitively between the encoder, trying to reduce the dimensions, and the decoder, trying to reconstruct the reduced values, while the error is calculated over the difference. If linear activation functions are used, the method has a similar connotation to a PCA approach, and hence it is recommended to use activation functions such as Sigmoid and ReLU. This dimensionality method has been extensively used for QML models catering to image-based datasets and tends to assist said models in achieving questionably high accuracies. 

\subsection{t-distributed Stochastic Neighbor Embedding (t-SNE)}
t-SNE is an unsupervised non-linear dimensionality reduction technique for data exploration and visualizing high-dimensional data introduced in \cite{vanDerMaaten2008}. The goal of the t-SNE algorithm is to take a set of points in a high-dimensional space and find a faithful representation of this space in a lower-dimensional space, mostly a 2D space for visualization. Another aspect of the t-SNE algorithm is the tunable parameter 'perplexity', which can be used to balance the local and global aspects of the data \cite{wattenberg2016how}.

\section{QML models}
\label{QML models}
QML models are generally a set of variational parameterized gate-based circuits \cite{bowles2024betterclassicalsubtleart}. These gates are generally rotational gates where the rotational parameters are the training parameters used for configuring the model based on some classical optimizers. While there has been considerable research and new ideas have been introduced for implementing quantum machine learning models, all these ideas can be easily differentiated into two types of variational models, i.e., Quantum Neural Networks and Quantum Kernel models. Even if \cite{bowles2024betterclassicalsubtleart} has differentiated Quantum Convolutional Neural Networks as an additional type during their implementation, they have structural commonalities with the QNN models with a few additional steps. Hence, for our experimentation, we have considered them under the QNN part.

Usually, classical data is added to these models using Data embedding methods, which use varying procedures to convert the classical information onto the generated QML model circuit. For our experimentation, we have used a combination of various data embedding techniques like Angle Embedding (on all three Pauli bases X, Y, and Z), Amplitude Embedding, and Instantaneous Quantum Polynomial (IQP) Embedding on all created QML models. Results of all these models have been presented in the section \ref{Results}. 

In the following subsection, we detail the general construction of these QML models and the manner in which we have built them to represent the generalized framework of these model types.

\subsection{Quantum Neural Networks models(QNN)}
Despite the use of the word Neural Networks, apart from a few conceptual correlations, QNNs are no more than variational circuits whose parameters are optimized using classical optimization methods such as Adam, Gradient Descent, or Nesterov Momentum. The conceptual correlation is evident in the fact that the input layer serves as the information encoding layer, the output layer as the layer for obtaining predicted values, and the hidden layers as random, parameterized rotational gates. These QNN models are extensively studied in the Quantum Machine Learning research. They could be simply represented as $f(\theta, x) = \mathrm{Tr}[O (x, \theta) \rho(x,\theta)]$, where $x$ are the input data points, $O$ represents observable parameters, $\theta$ is the trainable parameter. Quantum Neural Networks usually consist of several such combinations of variational circuits for classifying data. For training these models, a differentiable cost function is defined as $ \textit{L} (\theta, X, y) = \frac{1}{N}\sum_i \textit{l}(\theta, x_i, y_i)$, where the loss $\textit{l}(\theta, x_i, y_i)$ measures the model performance for a specific batch of training data points $x_i$ and true label $y_i$ we know, and X, y summaries all training inputs and labels into a matrix or a vector \cite{bowles2024betterclassicalsubtleart}.

\begin{figure}
    \centering
    \includegraphics[width=1.0\linewidth]{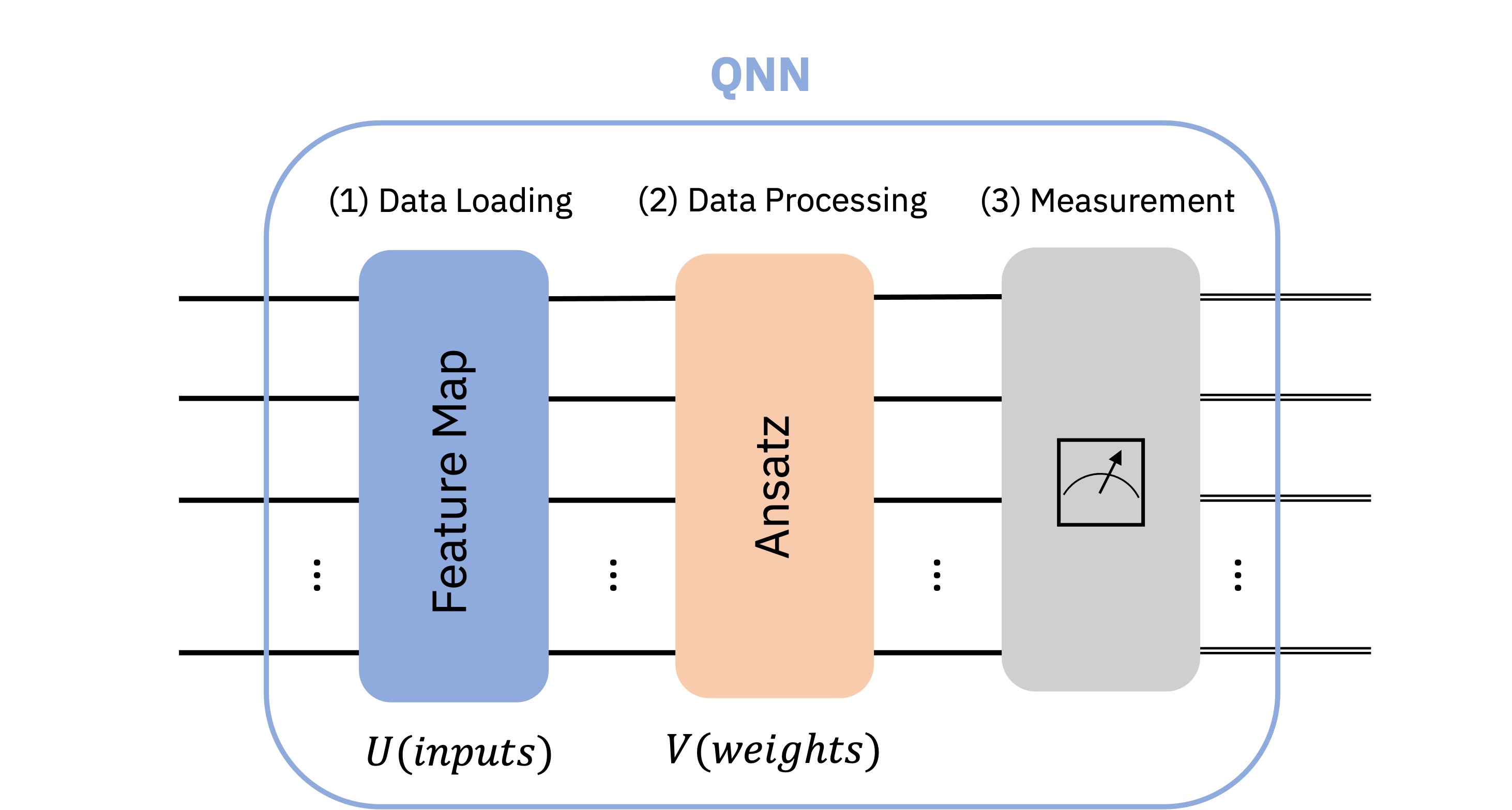}
    \caption{Generic quantum neural network (QNN) structure \cite{QNNQiskit}.}
    \label{fig:QNN}
\end{figure}

For our experimentation with the QNN models, we created two-qubit entangling circuits (Ansatz) inspired by the research presented in \cite{Hur_2022} for binary classification problems. This two-qubit entangling circuit was repeated for multiple layers and qubits to create a full quantum circuit, and the probabilities were measured on qubit '0' Fig. \ref{fig:QNN}. Hur et al. \cite{Hur_2022} presents a plethora of different Quantum ansatz with varying parameters and depths and has been proven to have considerably good performance \cite{ShindeOverfittingQCNN2024}. This helped us in verifying different QNN ansatz consisting of a total number of parameters ranging from 2 to 15 and with different entangling options. Another reason for choosing the 2-qubit ansatz is to create the structural parallelism for qubit models ranging from 16 qubits to 8 qubits. Ansatzes discussed in this manuscript have proven effective and complex enough to represent several difficult problems with considerable accuracy \cite{Hur_2022}\cite{ShindeOverfittingQCNN2024}. More about the structural configuration and reasoning will be discussed in Section \ref{Experimental Setup}.

\subsection{Quantum Kernel models}
In classical Machine Learning, Kernel methods have been extensively used and have been proven to be effective in solving problems related to the subdivision of hyperplanes. These methods could be represented as $f(x) = \sum_i \alpha_i k(x_i,x)$,  where $\alpha_i$ is the real trainable parameters, and $k$ represents a kernel function (positive definite function) quantifying similarities between the aforementioned data points $(x_i,x)$. A fundamental result in kernel methods states that these models correspond to a linear classifier operating in a potentially infinite-dimensional feature space \(|\phi(x)\rangle\), where the kernel function represents the inner product in this space:  $k(x, x') = \langle \phi(x) | \phi(x') \rangle$. In the case of Quantum Kernel methods, this kernel function $k$ is computed using a quantum computational method or quantum simulation. In most cases, for such implementation, one encodes data into a quantum state defined as $\rho(x)$ such that the resulting kernel is defined as $k(x_i,x_j)=tr[\rho(x_i)\rho(x_j)]$. This kernel can be further converted into a trainable option by using some classical optimization for the defined non-data parameters \cite{bowles2024betterclassicalsubtleart}. 

\begin{figure}
    \centering
    \includegraphics[width=0.80\linewidth]{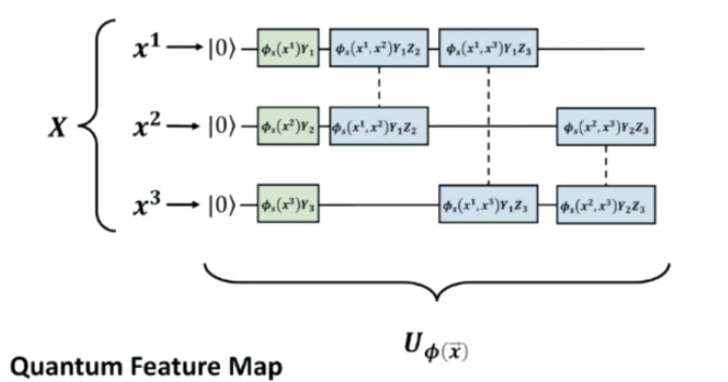}
    \includegraphics[width=0.80\linewidth]{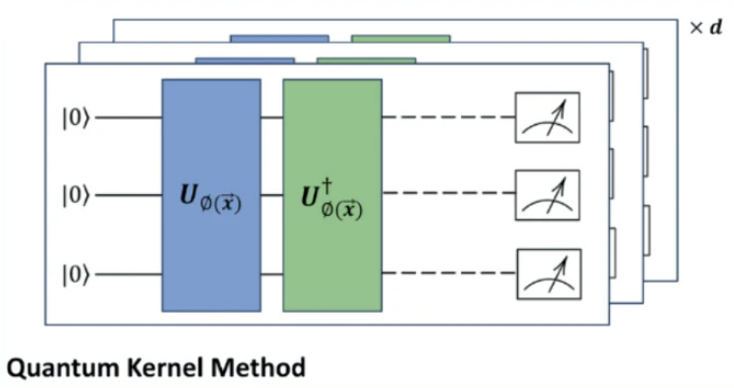}    
    \caption{Quantum SVM Kernel construction \cite{chen_quantum-enhanced_2024}}
    \label{fig:QKM}
\end{figure}

For our implementation of QSVC (Quantum Support Vector Classifier), a Quantum Kernel method, we derived our inspiration from the approach presented in \cite{schuld2021supervisedquantummachinelearning}\cite{Hubregtsen_2022} and \cite{Zhangmulti‐kernel}. We have constructed several different kernels with variable training parameters, number of qubits, and varying numbers of layers. These parameters are evaluated based on the target alignment method mentioned in \cite{Hubregtsen_2022} and optimized using classical optimization methods. Further, this trained kernel is provided to the SVC algorithm provided by Scikit-Learn \cite{scikitSVC} for fitting the training and predicting the test data values. We have also evaluated the models on the premise of the SVCLoss (Kernel Loss) function presented in \cite{Glick_2024}. While creating the kernel structure, we have also adhered to the caution presented in \cite{SalmenperäEmbeddingPlacement2024} as we were able to see the effects while experimenting.

\section{Generated Datasets Description}
\label{Dataset}
This experimentation was conducted on several datasets for binary classification. The datasets were generated using open-source algorithms and created a new dataset mimicking an image dataset for binary classification. Most real-world datasets are either too large for this experiment or have significant discrepancies for supporting binary classification; therefore, the initial phase of experimentation was conducted using the generated dataset. Fortunately, many more variations could be generated using the generated algorithms to get a varied assessment and prove the idea that the data reduction methods distill the information for QML models. Another setback with real-world image datasets is that no image dataset is available that can fit the limited qubits for simulation or isn't reduced using some manner to run small-scale ML models, hence the idea of creating representative datasets baring similarity with the image dataset with only 16 pixels altogether ($4 \times 4$). Bowles et al.\cite{bowles2024betterclassicalsubtleart} refer to several datasets generated for benchmarking QML models, but the complexity of these datasets wasn't adjustable and was limited to conveying our idea. Also, a few datasets were created using certain data reduction methods, countering the ideas that needed to be explored.

The following ways were used to generate data values to validate our findings:
\subsection{Make Classification by Scikit-Learn (Linear Dataset Creator)}
The algorithm used for generating datasets in the Make Classification function is based on the idea proposed in \cite{Guyon2003DesignOE} and was designed to generate the “Madelon” dataset. The 'Make Classification' function could be utilized for creating multiclass classification datasets, but due to the constraints put on the experimentation, we used it for creating Binary Class Classification datasets. The 'Make Classification' function excels in adding noise by attributes such as correlated, redundant, and uninformative features. It also adds noise by creating multiple Gaussian clusters per class and linear transformation of the feature space \cite{scikitSampleGenerators}.

\subsection{Synthetic Data by Capital One (Non-linear Dataset)}
The Synthetic Data generator is a tabular dataset creator that introduces non-linearity to the datasets, creating arbitrarily complex datasets. The major reason for choosing this dataset generator is to alleviate the concerns that data reduction techniques like PCA are favored by the simplistic Gaussian dataset generator and provide a complex dataset to verify our claims. Data reduction methods like PCA are good at condensing information in Linear Gaussian datasets, yet are more likely to underperform when facing the non-linear dataset created for deep learning models. 

This generation method employs a method such that it provides a joint probability $ P(X, y)$, where $X$ is the feature and $y$ refers to labels associated with $X$. $X$ is composed of three feature types: informative, redundant, and nuisance features. Informative features are generated in the determination of the binary labels and are specified by Copula theory, which is a mathematical framework for the separation of the correlation structure from the marginal distributions of the feature vectors. Whereas redundant features are some linear combination of informative features, and nuisance features are uncorrelated vectors drawn from the interval [-1,1]. More information about the dataset can be found here \cite{Barr2020TowardsGT}.

\subsection{ $4\times4$ Image Dataset}
A considerable number of Quantum Neural Network (QNN) models have been proposed for image classification problems. An overused example is the reduced MNIST dataset that used different data reduction methods and considerable preprocessing, catering to the limited number of available qubits. There are no image-based datasets with pixels limited enough to be naturally fitted into the QML model. Hence, we created a pseudo-image dataset that has 2 distinct objects labeled as 'diagonally aligned' and 'linearly aligned'.

\begin{figure}
  \centering
  \includegraphics[width=0.35\textwidth]{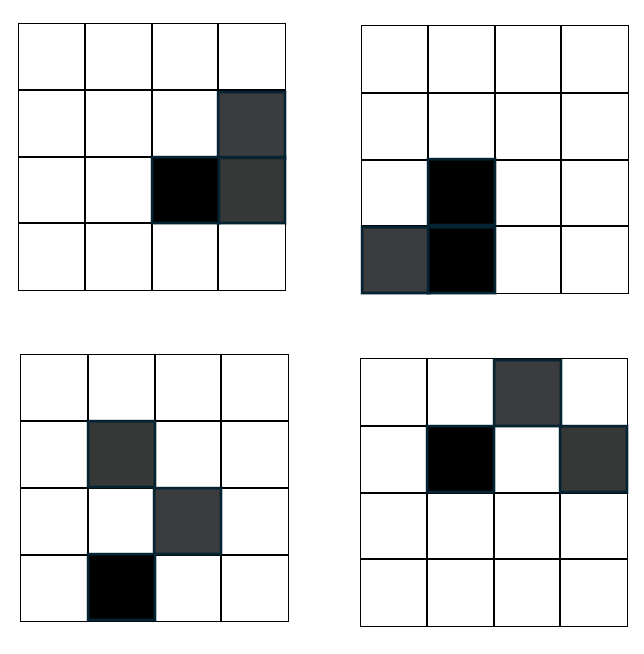}
  \caption{Example image of $4\times4$ Image Dataset describing two different classes termed as diagonally aligned or linearly aligned. }
  \label{fig:4x4_img}
\end{figure}

We have a few fair assumptions in the construction of the dataset regarding the usual preprocessing, such as gray scaling and flattening, which is consistent with all Image-based ML algorithms. Hence, the dataset is created with pre-normalized grayscale intensity pixel values and a 1-D array format. 
This dataset constructs two types of structures as mentioned in the Fig. \ref{fig:4x4_img}, labeled as 'diagonally aligned' and 'linearly aligned' with associated values of 0,1, respectively. Here, the pixel intensity varies to add variance to the data, and the intensity difference could be controlled during the dataset generation. Here, the high-intensity values are used to construct the structure. Also, the dataset could be created with an offset or balanced to add to the dataset complexity based on the percentage split.

\section{Experimental Setup}
\label{Experimental Setup}
To verify the effects of data reduction on the QML models, we first started with generating datasets using the methods described in Section \ref{Dataset}. The maximum number of features was equal to the maximum number of qubit models efficiently simulable on the local classical device, which in the case of QNN was 16 features and for QSVC was 8 features. Except for the $4\times4$ Image Dataset for the QSVC method, all the datasets were implemented for these models. The reason to do so is the constraints of fitting the $4\times4$ image onto an 8-qubit Quantum Kernel without data reduction for all embedding techniques and the classical processing constraints for a 16-qubit Quantum Kernel model. The parameters for all the dataset generation were kept constant over all the simulated models. To address the Randomizer problem in dataset generation, even the 'random state' parameter was fixed to a given value. Generated datasets were balanced for both the label values to remove any data value bias. All the datasets are divided into a Train-test split of 80-20, where the randomizer has been set to a random state. Several Embedding methods were implemented to observe the effects and the correlation of data-reduction methods with the embeddings. 

As the QNN model consisted of a 2-qubit recurring ansatz, the structure could be kept consistent; hence, we created QNN models ranging from 16 to 8 qubits with a reduction of 2 qubits each, resulting in 5 models for each type of ansatz (i.e. 16,14,12,10,8 qubit models). For the case of QSVC, we consider working with 8-qubit kernels with a reduction of 1 qubit in the following model, resulting in 5 models for different kernel structures (i.e. 8,7,6,5,4 qubit models).  

Apart from these QNN models, we have also experimented with a simple classical Convolutional Neural Network to showcase that there are no unimportant features in the dataset that had to be reduced. This Convolutional model consisted of two 1-D Convolutional Layers, a Pooling layer, and a Sigmoid Activation Function, which equated to a total of 1665 trainable parameters. 

Later, Data Reduction methods mentioned in Section \ref{DataRedu} were used on the generated dataset to fit the models generated using the aforementioned method. Variables for these methods were strictly fixed for all the models. Later, the data-reduced values fitted models and were compared against the model without the data-reduction techniques applied. We also used data reduction in the same dimensional space representation to observe the information distillation in the Quantum Models. This refers to applying a 0\% feature reduction, which is converting from 16 to 16 features for QNN and 8 to 8 features for Q-SVC. This also provides insights into the inherent effect of the general procedure of data dimensionality reduction methods and values generated by such methods on QML models. 

During the training and optimization procedure, variables like the 'batch size',' learning rate', and ' iterations' were kept unchanged for all the models. Also, the Optimization Function and Cost Function were kept constant while evaluating these models. In order to alleviate the concerns over the limited number of iterations for large models having more parameters to optimize and the convergence of the cost function being slow, we implemented a no data-reduction method for our largest model, for iterations 4 times the regular number of iterations. We observed no difference in convergence compared to the regular number of iterations in the same model with the no data-reduction method applied. All of these models are evaluated based on performance metrics of accuracy, precision, recall, and F1 score. 

The simulations were implemented using the Pennylane Framework \cite{pennylane2018arXiv181104968B} and classical simulation of quantum circuits to provide a noise-free environment for experimentation. Embedding methods were also implemented using the Pennylane library. Data generation was done using the source code mentioned in the description in Section \ref{Dataset}. For Convolutional Neural Network implementation Pytorch library was used. Similarly, Most of the data-reduction methods were implemented using the Scikit-Learn Library, except the autoencoder scheme, which was implemented using the TensorFlow open-source library. These results are generated over classical simulations done on a local device and the 'Puhti' Supercomputer.

\section{Results}
\label{Results}
A large number of simulations were performed with various combinations of variables such as ansatz, embedding methods, total number of qubits, optimization functions, and training iterations. The following results represent multiple different aspects of several experiments that were performed. These results have provided a consistent narrative with the other results and are a good depiction of the multitude of generated results. Tables \ref{tab:qnn_mk_class_data_amplitude}-\ref{tab:qnn_image} present results for Quantum Neural Network models, while tables \ref{tab:sklearn_cnn_table}-\ref{tab:4x4_cnn_table} display outcomes for the Classical Convolutional Neural Network model. Additionally, Tables \ref{tab:ckqk_class_sklearn} and \ref{tab:ckqk_class_cap1} showcase results for Quantum Kernel Methods.

Most of the tables have common column labels. Columns labeled 'Redu. $\Rightarrow$ 'Qbit no.' refers to the Reduction method used to reduce the features. In the case of Quantum Neural Network models, the features are reduced from 16 features, while in the case of Quantum Kernel models, the features are reduced from 8 features to the number of features denoted after the '$\Rightarrow$'. Other columns such as 'Acc.',' Prec.',' F1', 'Recall' refers to the evaluation metrics as accuracy, precision, F1, and recall. Apart from this, there are other columns such as 'Ansatz' and 'Embedding' referring to the Quantum Neural Network model ansatz and the classical to quantum data embedding method, respectively. The structure of the Ansatzes mentioned in these tables is depicted in the Appendix.

Table \ref{tab:qnn_mk_class_data_amplitude} presents the evaluations of different Quantum Neural Network ansatz on the premise of a common embedding method. The Amplitude embedding method was used for these results, and the linear datasets generation method was used for evaluations. Table \ref{tab:qnn_mk_class_SU4} provides a different perspective on the Quantum Neural Network experimentation by fixing the data reduction method and the QNN model but varying the data embedding method. This provides insight into how data reduction methods are dependent on the construction of the models and also on the embedding methods. Table \ref{tab:qnn_syn_dataset} refers to the results observed during the evaluation of the QNN models using Amplitude embedding on the non-linear Dataset. Table \ref{tab:qnn_image} presents data associated with the $4\times4$ Image Dataset, where the Angle Embedding with rotation along the 'X' axis is used for embedding the classical data. Table \ref{tab:sklearn_cnn_table}, \ref{tab:synthetic_cnn_table}, and \ref{tab:4x4_cnn_table} refer to the results of the simple CNN model for the Make Classification dataset by Sklearn, the Synthetic Dataset by Capital One, and the $4 \times 4$ Image datasets, respectively. Table \ref{tab:ckqk_class_sklearn} and Table \ref{tab:ckqk_class_cap1} showcase the performance of Classical and Quantum SVC models under the influence of data dimensionality reduction methods for linear and non-linear datasets, respectively. 

Values obtained by running QNN models without a data dimensionality reduction procedure had a mean accuracy of 0.5563 and a median accuracy of 0.545. The maximum accuracy observed was 0.74, with a minimum accuracy observed being 0.45, giving a range difference of 0.29. In all QNN model experimentation, we have seen the performance metrics of the QNN models with data reduction procedure excelling from those without. The maximum difference in accuracy observed was around +0.48, while the minimum difference was +0.25. The range of difference between these accuracies varies on the basis of datasets, methods of embedding, and ansatzes.

Table \ref{tab:ckqk_class_sklearn} and Table \ref{tab:ckqk_class_cap1} present a comparative study of the classical Support Vector Classification method and the Quantum Kernel-based Support Vector Classification method. These two tables showcase how data reduction methods affect the classical algorithms themselves, pertaining to the degradation of overall results for Quantum Methods.

For the Q-SVC cases, we have seen the same kind of loss in performance as in the Classical SVC case when data dimensionality reduction was introduced to these algorithms. The performance of all SVC models tends to reduce with the progressive reduction of the feature dimensions. Moreover, the loss of percentage in Q-SVC's performance has parallels with the loss in performance in Classical SVC. Overall, the difference between the performance values of classical and quantum SVC techniques does not vary more than $\pm0.05$, in association with the same reduction method and dataset.

All the observed data in the tables were generated over an average run of 10 values, with minimum runs being of 5 and maximum of 25, depending on the practical implementation phase of the experimentation. The values presented in these tables have a variance of $\pm0.02$. The best representative data is presented in the tables below. All of the generated data is available on GitHub and updated consistently with more runs.

\begin{table}[h]
        \centering
        \begin{tabular}{|c|c|c|c|c|c|} 
            \hline
            \textbf{Redu. $\Rightarrow$ dimen.} & \textbf{Ansatz} & \textbf{Acc.} & \textbf{Prec.} & \textbf{F1} & \textbf{Recall} \\
            \hline
            No Redu $\Rightarrow$ 16 &  \multirow{6}{*}{\parbox{1cm}{\centering U\_SU4 (15 params)}} & 0.491 & 1.0 & 0.659 & 0.491 \\ 
            PCA $\Rightarrow$ 16 & & 0.491 & 1.0 & 0.692 & 0.491 \\
            PCA $\Rightarrow$ 14 & & 0.983 & 1.0 & 0.93 & 0.967 \\
            PCA $\Rightarrow$ 12 & & 0.741 & 1.0 & 0.791 & 0.655 \\
            PCA $\Rightarrow$ 10 & & 1.0 & 1.0 & 1.0 & 1.0 \\
            PCA$\Rightarrow$ 8  & & 1.0 & 1.0 & 1.0 & 1.0 \\
            \hline
            No Redu $\Rightarrow$ 16 &  \multirow{6}{*}{\parbox{1cm}{\centering U\_14 (6 params)}} & 0.49 & 1.0 & 0.66 & 0.49 \\ 
            SVD $\Rightarrow$ 16 & & 0.47 & 1.0 & 0.63 & 0.47 \\
            SVD $\Rightarrow$ 14 & & 0.47 & 1.0 & 0.63 & 0.47 \\
            SVD $\Rightarrow$ 12 & & 0.54 & 1.0 & 0.66 & 0.5 \\
            SVD $\Rightarrow$ 10 & & 0.83 & 0.99 & 0.85 & 0.74 \\
            SVD $\Rightarrow$ 8  & & 0.90 & 0.90 & 0.89 & 0.88 \\
            \hline
            No Redu $\Rightarrow$ 16 &  \multirow{6}{*}{\parbox{1cm}{\centering U\_13 (6 params)}} & 0.55 & 0.0 & 0.6 & 0.0 \\ 
            Autoencode $\Rightarrow$ 16 & & 0.64 & 0.0 & 0.0 & 0.0 \\
            Autoencode $\Rightarrow$ 14 & & 0.54 & 0.0 & 0.0 & 0.0 \\
            Autoencode $\Rightarrow$ 12 & & 0.85 & 0.80 & 0.84 & 0.87 \\
            Autoencode $\Rightarrow$ 10 & & 0.89 & 0.9 & 0.87 & 0.84 \\
            Autoencode $\Rightarrow$ 8  & & 0.9 & 0.89 & 0.89 & 0.89 \\
            \hline
            No Redu $\Rightarrow$ 16 &  \multirow{6}{*}{\parbox{1cm}{\centering U\_9\_1D (2 params, no pooling)}} & 0.55 & 0.7796 & 0.630 & 0.5287\\ 
            PCA $\Rightarrow$ 16 & & 0.46 & 0.949 & 0.6363 & 0.4786 \\
            PCA $\Rightarrow$ 14 & & 0.76 & 0.5254 & 0.688 & 1.0 \\
            PCA $\Rightarrow$ 12 & & 0.933 & 0.9491 & 0.9333 & 0.9180 \\
            PCA $\Rightarrow$ 10 & & 0.9583 & 0.9661 & 0.957 & 0.95 \\
            PCA$\Rightarrow$ 8  & & 0.933 & 0.93 & 0.932 & 0.934 \\
          \hline
            No Redu $\Rightarrow$ 16 &  \multirow{6}{*}{\parbox{1cm}{\centering U\_5 (10 params)}} & 0.55 & 1.0 & 0.66 & 0.5 \\ 
            PCA $\Rightarrow$ 16 & & 1.0 & 1.0 & 1.0 & 1.0 \\
            PCA $\Rightarrow$ 14 & & 0.49 & 1.0 & 0.65 & 0.49 \\
            PCA $\Rightarrow$ 12 & & 1.0 & 1.0 & 1.0 & 1.0 \\
            PCA $\Rightarrow$ 10 & & 0.97 & 1.0 & 0.975 & 0.95 \\
            PCA $\Rightarrow$ 8  & & 0.9916 & 1.0 & 0.9915 & 0.981 \\
            \hline
            No Redu $\Rightarrow$ 16 &  \multirow{6}{*}{\parbox{1cm}{\centering U\_SU4 (15 params)}} & 0.47 & 1.0 & 0.63 & 0.47 \\ 
            SVD $\Rightarrow$ 16 & & 0.47 & 1.0 & 0.63 & 0.47 \\
            SVD $\Rightarrow$ 14 & & 0.47 & 1.0 & 0.63 & 0.47 \\
            SVD $\Rightarrow$ 12 & & 0.77 & 0.99 & 0.8 & 0.67 \\
            SVD $\Rightarrow$ 10 & & 0.57 & 0.06 & 0.12 & 1.00 \\
            SVD $\Rightarrow$ 8  & & 0.89 & 0.77 & 0.86 & 0.97 \\
            \hline
        \end{tabular}
        \vspace{3pt}
        \caption{\centering Quantum Neural Network Results: Make Classification Sklearn Dataset (Linear) with Amplitude Embedding}
        \label{tab:qnn_mk_class_data_amplitude}
        \vspace{2pt}
        
\end{table}

\begin{table}[h]
        \centering
        \begin{tabular}{|c|c|c|c|c|c|} 
            \hline
            \textbf{Redu. $\Rightarrow$ dimen.} & \textbf{Embedding} & \textbf{Acc.} & \textbf{Prec.} & \textbf{F1} & \textbf{Recall} \\
            \hline
            No Redu $\Rightarrow$ 16 &  \multirow{6}{*}{\parbox{1cm}{\centering Amplitude}} & 0.51 & 0.0 & 0.0 & 0.0 \\ 
            Autoencode $\Rightarrow$ 16 & & 0.49 & 1.0 & 0.66 & 0.49 \\
            Autoencode $\Rightarrow$ 14 & & 0.91 & 0.99 & 0.91 & 0.84 \\
            Autoencode $\Rightarrow$ 12 & & 0.51 & 1.0 & 0.67 & 0.50 \\
            Autoencode $\Rightarrow$ 10 & & 1.0 & 0.99 & 0.99 & 1.0 \\
            Autoencode$\Rightarrow$ 8  & & 0.99 & 0.99 & 0.99 & 0.98 \\
            \hline
            No Redu $\Rightarrow$ 16 &  \multirow{6}{*}{\parbox{1cm}{\centering Angle X}} & 0.69 & 0.8 & 0.7 & 0.63 \\ 
            Autoencode $\Rightarrow$ 16 & & 0.96 & 0.93 & 0.96 & 0.99 \\
            Autoencode $\Rightarrow$ 14 & & 0.99 & 0.98 & 0.99 & 0.99 \\
            Autoencode $\Rightarrow$ 12 & & 0.96 & 0.92 & 0.95 & 0.99 \\
            Autoencode $\Rightarrow$ 10 & & 0.99 & 0.97 & 0.98 & 0.99 \\
            Autoencode $\Rightarrow$ 8  & & 0.81 & 0.72 & 0.79 & 0.86 \\
            \hline
            No Redu $\Rightarrow$ 16 &  \multirow{6}{*}{\parbox{1cm}{\centering Angle Y}} & 0.67 & 0.8 & 0.7 & 0.63 \\ 
            Autoencode $\Rightarrow$ 16 & & 0.99 & 0.97 & 0.98 & 1.0 \\
            Autoencode $\Rightarrow$ 14 & & 0.99 & 0.98 & 0.99 & 0.99 \\
            Autoencode $\Rightarrow$ 12 & & 0.99 & 0.99 & 0.98 & 0.98 \\
            Autoencode $\Rightarrow$ 10 & & 0.95 & 0.90 & 0.95 & 1.0 \\
            Autoencode $\Rightarrow$ 8  & & 0.92 & 0.85 & 0.92 & 0.99 \\
            \hline
            No Redu $\Rightarrow$ 16 &  \multirow{6}{*}{\parbox{1cm}{\centering Angle Z}} & 0.51 & 0.0 & 0.0 & 0.0 \\ 
            Autoencode $\Rightarrow$ 16 & & 0.51 & 0.0 & 0.0 & 0.0 \\
            Autoencode $\Rightarrow$ 14 & & 0.49 & 1.0 & 0.66 & 0.49 \\
            Autoencode $\Rightarrow$ 12 & & 0.51 & 0.0 & 0.0 & 0.0 \\
            Autoencode $\Rightarrow$ 10 & & 0.49 & 1.0 & 0.66 & 0.49 \\
            Autoencode$\Rightarrow$ 8  & & 0.49 & 1.0 & 0.66 & 0.49 \\

            \hline
            No Redu $\Rightarrow$ 16 &  \multirow{6}{*}{\parbox{1cm}{\centering IQP}} & 0.59 & 0.58 & 0.58 & 0.58 \\ 
            Autoencode $\Rightarrow$ 16 & & 0.68 & 0.42 & 0.56 & 0.85 \\
            Autoencode $\Rightarrow$ 14 & & 0.84 & 0.67 & 0.8 & 1.0 \\
            Autoencode $\Rightarrow$ 12 & & 0.78 & 0.56 & 0.71 & 1.0 \\
            Autoencode $\Rightarrow$ 10 & & 0.73 & 0.57 & 0.67 & 0.82 \\
            Autoencode $\Rightarrow$ 8  & & 0.76 & 0.68 & 0.74 & 0.81 \\
            \hline
        \end{tabular}
        \vspace{3pt}
        \caption{\centering Quantum Neural Network Results: Make Classification Sklearn Dataset (Linear) for U\_SU4 model with different embedding and same dimensionality reduction method }
        \label{tab:qnn_mk_class_SU4}
        \vspace{2pt}
\end{table}

\begin{table}[h]
        \centering
        \begin{tabular}{|c|c|c|c|c|c|} 
            \hline
            \textbf{Redu. $\Rightarrow$ dimen.} & \textbf{Ansatz} & \textbf{Acc.} & \textbf{Prec.} & \textbf{F1} & \textbf{Recall} \\
            \hline
            No Redu $\Rightarrow$ 16 &  \multirow{6}{*}{\parbox{1cm}{\centering U\_TTN (2 params)}} & 0.51 & 0.0 & 0.0 & 0.0 \\ 
            TSNE $\Rightarrow$ 16 & & 0.51 & 0.0 & 0.0 & 0.0 \\ 
            TSNE $\Rightarrow$ 14 & & 0.51 & 0.0 & 0.0 & 0.0 \\ 
            TSNE $\Rightarrow$ 12 & & 0.49 & 1.0 & 0.6 & 0.49 \\ 
            TSNE $\Rightarrow$ 10 & & 0.80 & 0.60 & 0.75 & 0.98 \\
            TSNE $\Rightarrow$ 8  & & 0.88 & 0.93 & 0.88 & 0.84 \\
            \hline
            No Redu $\Rightarrow$ 16 &  \multirow{6}{*}{\parbox{1cm}{\centering U\_9\_1D (2 params, no pooling)}} & 0.45 & 1.0 & 0.62 & 0.45\\ 
            Autoencode $\Rightarrow$ 16 & & 0.53 & 0.02 & 0.03 & 0.20 \\
            Autoencode $\Rightarrow$ 14 & & 0.78 & 0.49 & 0.66 & 0.99 \\
            Autoencode $\Rightarrow$ 12 & & 0.90 & 0.86 & 0.88 & 0.90 \\
            Autoencode $\Rightarrow$ 10 & & 0.89 & 0.91 & 0.88 & 0.85 \\
            Autoencode $\Rightarrow$ 8  & & 0.85 & 0.76 & 0.81 & 0.88 \\
            \hline
            No Redu $\Rightarrow$ 16 &  \multirow{6}{*}{\parbox{1cm}{\centering U\_14 (6 params)}} & 0.54 & 0.0 & 0.0 & 0.0 \\ 
            Autoencode $\Rightarrow$ 16 & & 0.47 & 1.0 & 0.63 & 0.47 \\
            Autoencode $\Rightarrow$ 14 & & 0.78 & 0.61 & 0.72 & 0.88 \\
            Autoencode $\Rightarrow$ 12 & & 0.47 & 1.0 & 0.63 & 0.47 \\
            Autoencode $\Rightarrow$ 10 & & 0.90 & 0.85 & 0.89 & 0.93 \\
            Autoencode $\Rightarrow$ 8  & & 0.89 & 0.93 & 0.89 & 0.85 \\
            \hline
            No Redu $\Rightarrow$ 16 &  \multirow{6}{*}{\parbox{1cm}{\centering U\_9 (2 params)}} & 0.68 & 0.36 & 0.53 & 0.99 \\ 
            TSNE $\Rightarrow$ 16 & & 0.55 & 0.17 & 0.26 & 0.56 \\ 
            TSNE $\Rightarrow$ 14 & & 0.83 & 0.84 & 0.82 & 0.80 \\ 
            TSNE $\Rightarrow$ 12 & & 0.89 & 0.90 & 0.88 & 0.86 \\ 
            TSNE $\Rightarrow$ 10 & & 0.87 & 0.93 & 0.87 & 0.81 \\
            TSNE $\Rightarrow$ 8  & & 0.85 & 0.89 & 0.84 & 0.80 \\
            
            \hline
            No Redu $\Rightarrow$ 16 &  \multirow{6}{*}{\parbox{1cm}{\centering U\_SU4 (15 params, no pooling)}} & 0.47 & 1.0 & 0.63 & 0.47 \\ 
            TSNE $\Rightarrow$ 16 & & 0.47 & 1.0 & 0.63 & 0.47 \\ 
            TSNE $\Rightarrow$ 14 & & 0.47 & 1.0 & 0.63 & 0.47 \\
            TSNE $\Rightarrow$ 12 & & 0.89 & 0.85 & 0.87 & 0.9 \\ 
            TSNE $\Rightarrow$ 10 & & 0.88 & 0.85 & 0.87 & 0.89 \\
            TSNE $\Rightarrow$ 8  & & 0.88 & 0.9 & 0.88 & 0.85 \\
            \hline
        \end{tabular}
        \vspace{3pt}
        \caption{\centering Quantum Neural Network Results: Synthetic Data, Capitol One (Non-Linear) Dataset with Amplitude Embedding}
        \label{tab:qnn_syn_dataset}
        \vspace{2pt}
        
\end{table}

\begin{table}[h]
        \centering
        
        \begin{tabular}{|c|c|c|c|c|c|} 
            \hline
            \textbf{Redu. $\Rightarrow$ dimen.} & \textbf{Ansatz} & \textbf{Acc.} & \textbf{Prec.} & \textbf{F1} & \textbf{Recall} \\
            \hline
            No Redu $\Rightarrow$ 16 &  \multirow{6}{*}{\parbox{1cm}{\centering U\_13 (6 params)}} & 0.54 & 1.0 & 0.68 & 0.52 \\ 
            Autoencode $\Rightarrow$ 16 & & 0.80 & 0.77 & 0.79 & 0.82 \\ 
            Autoencode $\Rightarrow$ 14 & & 0.76 & 0.76 & 0.75 & 0.73 \\ 
            Autoencode $\Rightarrow$ 12 & & 0.78 & 0.86 & 0.8 & 0.74 \\ 
            Autoencode $\Rightarrow$ 10 & & 0.87 & 0.84 & 0.87 & 0.90 \\
            Autoencode $\Rightarrow$ 8  & & 0.48 & 0.0 & 0.0 & 0.0 \\
            \hline
            No Redu $\Rightarrow$ 16 &  \multirow{6}{*}{\parbox{1cm}{\centering U\_6 (10 params)}} & 0.58 & 0.87 & 0.68 & 0.55 \\ 
            Autoencode $\Rightarrow$ 16 & & 0.76 & 0.72 & 0.75 & 0.79 \\ 
            Autoencode $\Rightarrow$ 14 & & 0.87 & 0.82 & 0.86 & 0.90 \\ 
            Autoencode $\Rightarrow$ 12 & & 0.79 & 0.74 & 0.78 & 0.89 \\ 
            Autoencode $\Rightarrow$ 10 & & 0.83 & 0.68 & 0.81 & 1.0 \\
            Autoencode $\Rightarrow$ 8  & & 0.82 & 0.83 & 0.82 & 0.80 \\
            \hline
            No Redu $\Rightarrow$ 16 &  \multirow{6}{*}{\parbox{1cm}{\centering U\_SO4 (15 params)}} & 0.49 & 1.0 & 0.65 & 0.48 \\ 
            Autoencode $\Rightarrow$ 16 & & 0.85 & 0.78 & 0.84 & 0.90 \\ 
            Autoencode $\Rightarrow$ 14 & & 0.90 & 0.83 & 0.89 & 0.96 \\ 
            Autoencode $\Rightarrow$ 12 & & 0.72 & 0.70 & 0.72 & 0.75 \\ 
            Autoencode $\Rightarrow$ 10 & & 0.76 & 0.70 & 0.74 & 0.79 \\
            Autoencode $\Rightarrow$ 8  & & 0.78 & 0.80 & 0.78 & 0.77 \\
            \hline
            No Redu $\Rightarrow$ 16 &  \multirow{6}{*}{\parbox{1cm}{\centering U\_SU4 \_1D (15 params)}} & 0.74 & 0.85 & 0.76 & 0.69 \\ 
            Autoencode $\Rightarrow$ 16 & & 0.88 & 0.81 & 0.87 & 0.93 \\ 
            Autoencode $\Rightarrow$ 14 & & 0.86 & 0.81 & 0.86 & 0.91 \\ 
            Autoencode $\Rightarrow$ 12 & & 0.87 & 0.87 & 0.87 & 0.86 \\ 
            Autoencode $\Rightarrow$ 10 & & 0.87 & 0.87 & 0.86 & 0.86 \\
            Autoencode $\Rightarrow$ 8  & & 0.86 & 0.85 & 0.86 & 0.88 \\
            \hline
            No Redu $\Rightarrow$ 16 &  \multirow{6}{*}{\parbox{1cm}{\centering U\_SU4 (15 params, no pooling)}} & 0.62 & 0.75 & 0.66 & 0.59 \\ 
            Autoencode $\Rightarrow$ 16 & & 0.77 & 0.67 & 0.74 & 0.82 \\ 
            Autoencode $\Rightarrow$ 14 & & 0.75 & 0.69 & 0.73 & 0.77 \\ 
            Autoencode $\Rightarrow$ 12 & & 0.85 & 0.83 & 0.85 & 0.87 \\ 
            Autoencode $\Rightarrow$ 10 & & 0.72 & 0.74 & 0.73 & 0.71 \\
            Autoencode $\Rightarrow$ 8  & & 0.89 & 0.79 & 0.88 & 0.99 \\
            \hline
        \end{tabular}
        \vspace{3pt}
        \caption{\centering Quantum Neural Network Results: $4\times4$ Image Dataset 'X' Rotation Angle Embedding}
        \label{tab:qnn_image}
        \vspace{2pt}
        
\end{table}

\begin{table}[h]{}
        \centering
        \begin{tabular}{|c|c|c|c|c|} 
            \hline
            \textbf{Redu. $\Rightarrow$ dimen.} & \textbf{Acc.} & \textbf{Prec.} & \textbf{F1} & \textbf{Recall} \\
            \hline
            No Redu $\Rightarrow$ 16 & 0.99 & 1.0 & 0.98 & 0.99 \\ 
            \hline
            PCA $\Rightarrow$ 16 & 0.9875 & 0.9897 & 0.9647 & 0.9872 \\ 
            PCA $\Rightarrow$ 14 & 0.9875 & 0.9897 & 0.9847 & 0.9872 \\ 
            PCA $\Rightarrow$ 12 & 0.9825 & 0.9892 & 0.9847 & 0.9622 \\ 
            PCA $\Rightarrow$ 10 & 0.9875 & 0.9846 & 0.9696 & 0.9621 \\
            PCA $\Rightarrow$ 8  & 0.9800 & 0.9947 & 0.9647 & 0.9797 \\
            \hline
            Autoencode $\Rightarrow$ 16 & 0.9900 & 0.9898 & 0.9898 & 0.9898 \\ 
            Autoencode $\Rightarrow$ 14 & 0.9950 & 1.0 & 0.9898 & 0.9949 \\ 
            Autoencode $\Rightarrow$ 12 & 0.9925 & 0.9949 & 0.9898 & 0.9923 \\ 
            Autoencode $\Rightarrow$ 10 & 0.9900 & 0.9898 & 0.9898 & 0.9898 \\
            Autoencode $\Rightarrow$ 8 & 0.9950 & 1.0 & 0.9898 & 0.9949 \\
            \hline
            SVD $\Rightarrow$ 16 & 0.9925 & 1.0 & 0.9847 & 0.9923 \\ 
            SVD $\Rightarrow$ 14 & 0.9900 & 0.9898 & 0.9898 & 0.9898 \\ 
            SVD $\Rightarrow$ 12 & 0.9900 & 0.9898 & 0.9898 & 0.9898 \\ 
            SVD $\Rightarrow$ 10 & 0.9825 & 0.9749 & 0.9898 & 0.9823 \\
            SVD $\Rightarrow$ 8  & 0.9925 & 0.9849 & 0.9798 & 0.9846 \\
            \hline
            TSNE $\Rightarrow$ 16 & 0.8530 & 0.8966 & 0.7959 & 0.8432 \\ 
            TSNE $\Rightarrow$ 14 & 0.8450 & 0.7617 & 0.9949 & 0.8628 \\ 
            TSNE $\Rightarrow$ 12 & 0.7550 & 0.9016 & 0.5612 & 0.6918 \\ 
            TSNE $\Rightarrow$ 10 & 0.9900 & 0.9898 & 0.9898 & 0.9898 \\
            TSNE $\Rightarrow$ 8  & 0.9600 & 0.9500 & 0.9690 & 0.9596 \\
            \hline
        \end{tabular}
        \vspace{3pt}
        \caption{Sklearn Make Class.: Classical Convolutional Neural Network Evaluations }
        \label{tab:sklearn_cnn_table}
        \vspace{2pt}
        
\end{table}
            
\begin{table}[h]{}
        \centering
        \begin{tabular}{|c|c|c|c|c|} 
            \hline
            \textbf{Redu. $\Rightarrow$ dimen.} & \textbf{Acc.} & \textbf{Prec.} & \textbf{F1} & \textbf{Recall} \\
            \hline
             No Redu $\Rightarrow$ 16  & 0.89 & 0.8817 & 0.8817 & 0.8817 \\
            \hline
            PCA $\Rightarrow$ 16 & 0.9 & 0.9011 & 0.8817 & 0.8913 \\ 
            PCA $\Rightarrow$ 14 & 0.8850 & 0.8500 & 0.9140 & 0.8808 \\ 
            PCA $\Rightarrow$ 12 & 0.872 & 0.8649 & 0.8602 & 0.8625 \\ 
            PCA $\Rightarrow$ 10 & 0.8850 & 0.8846 & 0.8656 & 0.8750 \\
            PCA $\Rightarrow$ 8  & 0.90 & 0.8842 & 0.9032 & 0.8936 \\
            \hline
            Autoencode $\Rightarrow$ 16 & 0.8875 & 0.8615 & 0.9032 & 0.8819 \\ 
            Autoencode $\Rightarrow$ 14 & 0.8925 & 0.8743 & 0.8978 & 0.8859 \\ 
            Autoencode $\Rightarrow$ 12 & 0.8875 & 0.8730 & 0.8871 & 0.8800 \\ 
            Autoencode $\Rightarrow$ 10 & 0.8900 & 0.8777 & 0.8871 & 0.8824 \\
            Autoencode $\Rightarrow$ 8 & 0.8925 & 0.8783 & 0.8925 & 0.8853 \\
            \hline
            SVD $\Rightarrow$ 16 & 0.8775 & 0.8703 & 0.8656 & 0.8679 \\ 
            SVD $\Rightarrow$ 14 & 0.8850 & 0.8684 & 0.8871 & 0.8777 \\ 
            SVD $\Rightarrow$ 12 & 0.8900 & 0.9034 & 0.8548 & 0.8785 \\ 
            SVD $\Rightarrow$ 10 & 0.9 & 0.8724 & 0.9194 & 0.8953 \\
            SVD $\Rightarrow$ 8 & 0.9050 & 0.8776 & 0.9247 & 0.9005 \\
            \hline
            TSNE $\Rightarrow$ 16 & 0.8525 & 0.8325 & 0.8548 & 0.8435 \\ 
            TSNE $\Rightarrow$ 14 & 0.7675 & 0.6824 & 0.9355 & 0.7891 \\ 
            TSNE $\Rightarrow$ 12 & 0.8375 & 0.7788 & 0.9086 & 0.8387 \\ 
            TSNE $\Rightarrow$ 10 & 0.7450 & 0.6489 & 0.9839 & 0.7821 \\
            TSNE $\Rightarrow$ 8  & 0.8425 & 0.8060 & 0.8710 & 0.8312 \\
            \hline
        \end{tabular}
        \vspace{3pt}
        \caption{Synthetic Data: Classical Convolutional Neural Network Evaluations }
        \label{tab:synthetic_cnn_table}
        \vspace{2pt}
        
\end{table}

\begin{table}[h]{}
        \centering
        \begin{tabular}{|c|c|c|c|c|} 
            \hline
            \textbf{Redu. $\Rightarrow$ dimen.} & \textbf{Acc.} & \textbf{Prec.} & \textbf{F1} & \textbf{Recall} \\
            \hline
            No Redu $\Rightarrow$ 16 & 1.0 & 1.0 & 1.0 & 1.0 \\ 
            \hline
            PCA $\Rightarrow$ 16 & 0.6210 & 0.5996 & 0.7611 & 0.6721 \\ 
            PCA $\Rightarrow$ 14 & 0.8710 & 0.9079 & 0.8362 & 0.8706 \\ 
            PCA $\Rightarrow$ 12 & 0.6330 & 0.6110 & 0.7793 & 0.6850 \\ 
            PCA $\Rightarrow$ 10 & 0.7330 & 0.6883 & 0.7894 & 0.7354 \\
            PCA $\Rightarrow$ 8  & 0.4780 & 0.4868 & 0.5404 & 0.5182 \\
            \hline
            Autoencode $\Rightarrow$ 16 & 0.9890 & 1.0 & 0.9783 & 0.9891 \\ 
            Autoencode $\Rightarrow$ 14 & 0.9050 & 0.8849 & 0.9324 & 0.9080 \\ 
            Autoencode $\Rightarrow$ 12 & 0.9530 & 0.9149 & 1.0 & 0.9554 \\ 
            Autoencode $\Rightarrow$ 10 & 0.9970 & 1.0 & 0.9940 & 0.9970 \\
            Autoencode $\Rightarrow$ 8 & 0.9220 & 0.9580 & 0.8872 & 0.9212 \\
            \hline 
            SVD $\Rightarrow$ 16 & 0.9070 & 0.9460 & 0.8530 & 0.8979 \\ 
            SVD $\Rightarrow$ 14 & 0.7640 & 0.7164 & 0.8692 & 0.7855 \\ 
            SVD $\Rightarrow$ 12 & 0.7810 & 0.7267 & 0.8880 & 0.7993 \\ 
            SVD $\Rightarrow$ 10 & 0.8430 & 0.8205 & 0.8691 & 0.8441 \\
            SVD $\Rightarrow$ 8  & 0.7820 & 0.7618 & 0.7996 & 0.7802 \\
            \hline
            TSNE $\Rightarrow$ 16 & 0.5350 & 0.5463 & 0.5833 & 0.5642 \\ 
            TSNE $\Rightarrow$ 14 & 0.6590 & 0.6482 & 0.6835 & 0.6654 \\ 
            TSNE $\Rightarrow$ 12 & 0.5970 & 0.6 & 0.7059 & 0.6486 \\ 
            TSNE $\Rightarrow$ 10 & 0.8040 & 0.7289 & 0.9680 & 0.8316 \\
            TSNE $\Rightarrow$ 8  & 0.5350 & 0.5263 & 0.5274 & 0.5268 \\
            \hline
        \end{tabular}
        \vspace{3pt}
        \caption{$4\times4$ Image Dataset: Classical Convolutional Neural Network Evaluations }
        \label{tab:4x4_cnn_table}
        \vspace{2pt}
        
\end{table}

\begin{table}[h]{}
        \centering
        \begin{tabular}{|c|c|c|c|c|c|} 
            \hline
            \textbf{Redu. $\Rightarrow$ dimen.} & \textbf{Algo. type} & \textbf{Acc.} & \textbf{Prec.} & \textbf{F1} & \textbf{Recall} \\
            \hline
            No Redu $\Rightarrow$ 8 & \multirow{6}{*}{\parbox{1cm}{\centering Q-SVC}} & 0.96 & 0.95 & 0.96 & 0.96 \\ 
            Autoencode $\Rightarrow$ 8 & & 0.92 & 0.93 & 0.92 & 0.91 \\ 
            Autoencode $\Rightarrow$ 7 & & 0.91 & 0.90 & 0.91 & 0.91 \\ 
            Autoencode $\Rightarrow$ 6 & & 0.90 & 0.88 & 0.89 & 0.90 \\ 
            Autoencode $\Rightarrow$ 5 & & 0.80 & 0.76 & 0.79 & 0.83 \\
            Autoencode $\Rightarrow$ 4  & & 0.66 & 0.68 & 0.67 & 0.60 \\
            \hline
            No Redu $\Rightarrow$ 8 & \multirow{6}{*}{\parbox{1cm}{\centering Classical SVC}} & 0.95 & 0.9688& 0.9507 & 0.9323 \\ 
            Autoencode $\Rightarrow$ 8 & & 0.8795 & 0.9346 & 0.8836 & 0.8378 \\ 
            Autoencode $\Rightarrow$ 7 & & 0.9175 & 0.9447 & 0.9193 & 0.8952 \\ 
            Autoencode $\Rightarrow$ 6 & & 0.85 & 0.7788 & 0.8378 & 0.9064 \\ 
            Autoencode $\Rightarrow$ 5 & & 0.81 & 0.7437 & 0.7956 & 0.8554 \\
            Autoencode $\Rightarrow$ 4  & & 0.7675 & 0.7688 & 0.7234 & 0.6830 \\
            \hline
            No Redu $\Rightarrow$ 8 & \multirow{6}{*}{\parbox{1cm}{\centering Q-SVC}} & 0.96 & 0.95 & 0.96 & 0.96 \\ 
            PCA $\Rightarrow$ 8 & & 0.70 & 0.71 & 0.70 & 0.69 \\ 
            PCA $\Rightarrow$ 7 & & 0.70 & 0.70 & 0.70 & 0.70 \\ 
            PCA $\Rightarrow$ 6 & & 0.70 & 0.71 & 0.70 & 0.69 \\ 
            PCA $\Rightarrow$ 5 & & 0.70 & 0.71 & 0.70 & 0.69 \\
            PCA $\Rightarrow$ 4  & & 0.70 & 0.70 & 0.70 & 0.70 \\
            \hline
            No Redu $\Rightarrow$ 8 & \multirow{6}{*}{\parbox{1cm}{\centering Classical SVC}} & 0.9825 & 0.9798& 0.9823 & 0.9848 \\ 
            PCA $\Rightarrow$ 8 & & 0.70 & 0.6783 & 0.6923 & 0.7068 \\ 
            PCA $\Rightarrow$ 7 & & 0.70 & 0.6783 & 0.6923 & 0.7068 \\
            PCA $\Rightarrow$ 6 & & 0.70 & 0.6783 & 0.6923 & 0.7068 \\
            PCA $\Rightarrow$ 5 & & 0.70 & 0.6783 & 0.6923 & 0.7068 \\
            PCA $\Rightarrow$ 4 & & 0.70 & 0.6783 & 0.6923 & 0.7068 \\
            \hline
        \end{tabular}
        \vspace{3pt}
        \caption{Kernel Classification: Make Classification Sklearn Dataset (Linear) }
        \label{tab:ckqk_class_sklearn}
        \vspace{2pt}
        
\end{table}

\begin{table}[h]{}
        \centering
        \begin{tabular}{|c|c|c|c|c|c|} 
            \hline
            \textbf{Redu. $\Rightarrow$ dimen.} & \textbf{Algo. type} & \textbf{Acc.} & \textbf{Prec.} & \textbf{F1} & \textbf{Recall} \\
            \hline
            No Redu $\Rightarrow$ 8 & \multirow{6}{*}{\parbox{1cm}{\centering Q-SVC}} & 0.97 & 0.98 & 0.97 & 0.96 \\ 
            Autoencode $\Rightarrow$ 8 & & 0.92 & 0.94 & 0.91 & 0.89 \\ 
            Autoencode $\Rightarrow$ 7 & & 0.91 & 0.94 & 0.91 & 0.88 \\ 
            Autoencode $\Rightarrow$ 6 & & 0.91 & 0.94 & 0.91 & 0.88 \\ 
            Autoencode $\Rightarrow$ 5 & & 0.82 & 0.74 & 0.79 & 0.85 \\
            Autoencode $\Rightarrow$ 4  & & 0.84 & 0.84 & 0.83 & 0.83 \\
            \hline
            No Redu $\Rightarrow$ 8 & \multirow{6}{*}{\parbox{1cm}{\centering Classical SVC}} & 0.9825 & 0.9798& 0.9823 & 0.9848 \\ 
            Autoencode $\Rightarrow$ 8 & & 0.9125 & 0.9396 & 0.9144 & 0.8904 \\ 
            Autoencode $\Rightarrow$ 7 & & 0.8925 & 0.9193 & 0.8948 & 0.8714 \\ 
            Autoencode $\Rightarrow$ 6 & & 0.8925 & 0.8994 & 0.8927 & 0.8861 \\ 
            Autoencode $\Rightarrow$ 5 & & 0.8875 & 0.8793 & 0.8860 & 0.8928 \\
            Autoencode $\Rightarrow$ 4  & & 0.8925 & 0.8793 & 0.8905 & 0.9020 \\
            \hline
            No Redu $\Rightarrow$ 8 & \multirow{6}{*}{\parbox{1cm}{\centering Q-SVC}} & 0.97 & 0.98 & 0.97 & 0.96\\ 
            PCA $\Rightarrow$ 8 & & 0.92 & 0.93 & 0.91 & 0.90 \\ 
            PCA $\Rightarrow$ 7 & & 0.91 & 0.92 & 0.90 & 0.88 \\ 
            PCA $\Rightarrow$ 6 & & 0.90 & 0.93 & 0.9 & 0.88 \\ 
            PCA $\Rightarrow$ 5 & & 0.87 & 0.90 & 0.87 & 0.84 \\
            PCA $\Rightarrow$ 4  & & 0.91 & 0.93 & 0.91 & 0.89 \\
            \hline
            No Redu $\Rightarrow$ 8 & \multirow{6}{*}{\parbox{1cm}{\centering Classical SVC}} & 0.9825 & 0.9798& 0.9823 & 0.9848 \\ 
            PCA $\Rightarrow$ 8 & & 0.8675 & 0.8793 & 0.8684 & 0.8578 \\ 
            PCA $\Rightarrow$ 7 & & 0.8675 & 0.8793 & 0.8684 & 0.8578 \\ 
            PCA $\Rightarrow$ 6 & & 0.85 & 0.8643 & 0.8514 & 0.8390 \\ 
            PCA $\Rightarrow$ 5 & & 0.8475 & 0.8592 & 0.8486 & 0.8382 \\
            PCA $\Rightarrow$ 4  & & 0.86 & 0.8793 & 0.8620 & 0.8454 \\
            \hline
        \end{tabular}
        \vspace{3pt}
        \caption{Kernel Classification: Synthetic Data, Capital One (Non-Linear) Dataset}
        \label{tab:ckqk_class_cap1}
        
\end{table}

\section{Discussions}
\label{Discuss}
 Overall summary of the data presented in the manuscript hovers over the conclusion that the data dimensionality reduction methods affect the quantum machine learning methods differently. In a few models, it extenuates the results and performance obtained from said models, while in other cases, it affects negatively due to the inherent flaws in the classical components associated with the quantum model. The efficacy of the quantum machine learning models depends on several other factors, along with data dimensionality methods such as the percentage of feature reduction, the procedure of integrating classical data onto a quantum system, and overall execution of the said quantum system.

Observing the results for the Quantum Neural Network models, it is easy to speculate that QNN models perform better with the aid of data reduction methods. Results in Table \ref{tab:qnn_mk_class_data_amplitude} consistently showcase that without the use of data reduction methods, the ansatzes associated with the models are unable to achieve respectable accuracy. 

In multiple observations, it has been seen that when there is a 0\% feature reduction implemented, the results seem to be consistent with the no dimensionality reduction implementation, as there is no knowledge distillation for the model to perform. This is not always true, as observed in Table \ref{tab:qnn_image}, that in a few cases, using 0\%  feature reduction methods results in better performance of the quantum neural network models. This leads to the conclusion that the QNN models rely heavily on data reduction methods to converge into a solution.

The majority of the time, it has been observed that as the information is condensed down to lower dimensions, the model tends to perform better, but this scenario is not always consistent. Sometimes, too much reduction of feature space results in the loss of important constraints associated with the data information. Hence, the overall effects of the percentage of dimensionality reduction affect different ansatzes differently due to the number of tunable parameters.

Table \ref{tab:qnn_mk_class_SU4} shows that the use of data dimensionality reduction methods contributes to an increase in the accuracies achievable by the Quantum Neural Network models, regardless of the embedding methods. We also observe that for the particular model presented in the table \ref{tab:qnn_mk_class_SU4}, Angle Embedding with Z-axis rotation does not contribute whatsoever to the improvement of the overall accuracy across different percentages of reduction. This is a problem associated with the ansatz and the embedding values being commutative and not having any significant impact on the training of the QML model. We were able to see the same embedding method perform better comparatively for other ansatzes and reduction methods. 
Observations in Table \ref{tab:qnn_mk_class_SU4} are consistent for both non-linear Datasets and the $4\times4$ image dataset. 

Table \ref{tab:qnn_syn_dataset} cements the conclusions drawn from Table \ref{tab:qnn_mk_class_data_amplitude} from the perspective of a non-linear dataset. It also provides important insights for choosing an appropriate data dimensionality reduction method for the particular type of dataset. The data dimensionality techniques 'Autoencode' and 'T-SNE' are particularly attuned to perform better than the other 'linear' methods. Using methods like PCA for such datasets resulted in a depletion of accuracy by a range of 0.05-0.10. Yet linear data dimensionality reduction methods outperformed no-reduction conditions. This drastically depends upon the noise associated with the data generation.

Table \ref{tab:qnn_image} is essential to show that this observed phenomenon extends to image-based datasets as well. This is an important result as most of the QNN algorithms tend to cement their validity by testing on datasets such as MNIST and tend to use data dimensionality reduction methods to fit the data to a lower qubit value. 

Another concern arises that does this phenomenon of data dimensionality reduction also extend to the classical CNN models? Is there a necessity for data dimensionality reduction inherent to the generated datasets? To address said issues, we ran the experiment on a simple Classical Convolutional Neural Network model for all the generated datasets, whose results are presented individually in Tables \ref{tab:sklearn_cnn_table}, \ref{tab:synthetic_cnn_table}, and \ref{tab:4x4_cnn_table}. Classical CNN observations indicate that data dimensionality reduction either has no significant impact on results or leads to adverse effects, depending on the method used. This aligns with the theory of potential information loss associated with dimensionality reduction \cite{van2009dimensionality}. The comparison of classical CNN and QNN models shows no significant difference in their best results across all datasets, indicating both models have similar complexity for binary classification.

From the overall observation of the quantum neural network models, we can conclude that, even though the data dimension reduction methods aid in better performance of QNN models, the results are inconsistent and suggest volatility in QNN models. Apart from that, further research is needed in regards to the symbiosis of data reduction methods, ansatz, and data embedding methods. 

Results for Quantum SVC reveal the opposite view to what we observed for the QNN implementations. As presented in Table \ref{tab:ckqk_class_sklearn}  and \ref{tab:ckqk_class_cap1}, if data dimensionality reduction is used on the SVC algorithms, the models seemingly underperform for the generated datasets versus the model not use data dimensionality reduction. We also observe that this performance loss does not just conform to Quantum-SVC but also to Classical-SVC. As discussed in Section \ref{QML models}, in Quantum SVC methods, only Kernels are the 'quantum' component. These kernels are provided to the Classical-SVC algorithm to perform the fitting and testing of data. Hence, this is an inherent flaw of the classical-SVC component, which extends into the Quantum-SVC implementation. 

Upon further experimentation on the classical SVC, we were able to decipher that when we increase the total number of data points, the accuracy difference between reduced and non-reduced methods decreases. To achieve such convergence of performance metrics, we needed data points 10 to 20 times the original number. This results in slowing the SVC implementation classically. Further, performing a similar evaluation for the quantum SVC is extremely strenuous on the classical simulation systems. Hence, considering the setbacks of the classical Support Vector Classification method and the limitations of classical simulations, the use of a data reduction method for Quantum Kernel methods would rarely result in consistent results and seemingly might underperform.

Despite conducting numerous simulations across a wide variety of ansatzes, qubit configurations, embedding methods, and data dimensionality techniques, we have not found definitive evidence identifying which method is compatible with a specific ansatz or the optimal percentage of reduction for achieving the best results. Additionally, we are unable to identify an ansatz that is minimally impacted by data dimensionality reduction. However, we can conclusively state that Quantum Neural Networks (QNNs) benefit from data dimensionality reduction methods, while Quantum Support Vector Machines (QSVMs) experience decreased performance due to their reliance on classical Support Vector Classifier (SVC) components. 

\section{Conclusion}

The usage of data dimensionality reduction methods for implementing quantum machine learning models would result in questionable performance regarding such models. Even though the results of Quantum Neural Network models signify that using data dimensionality reduction procedures could lead to improvement in the overall performance of the Quantum Neural Network models, the success of the performance cannot be completely attributed to the Quantum Neural Network models. Apart from that, there is the issue of scalability. Using data dimensionality reduction methods could result in slowing down the overall procedure of implementing effective Quantum Neural Network models. 

In the case of Quantum Kernel methods, the data dimensionality reduction results in the depletion of the Q-SVM performance. This can be attributed to the inherent flaws in the associated classical components. Circumventing this setback requires a larger dataset size, but this results in a slower Quantum Kernel implementation. This brings back the issue of the scalability of quantum algorithms.

Hence, we conclude that it is essential to develop and implement quantum machine learning methods that are less reliant on data reduction methods. Apart from this, we suggest reevaluation of Quantum Machine Learning methods that use data dimensionality reduction methodology and claim a quantum advantage in performance over classical Machine Learning. 

\vspace{5pt}
\begin{center}
    \centering{ACKNOWLEDGMENT}
\end{center}

Special thanks to Ilmo Salmenperä and Valter Uotila for their valuable comments on the article's readability and discussion on the fundamentals of implementing the experiments.

\vspace{5pt}
\begin{center}
    \centering{DATA AVAILABILITY}
\end{center}

The source code used in this study and the generated results are available on the following GitHub links. 
\newline
QNN: \url{https://github.com/AakashShindeHelsinki/Data_Reduction_and_QCNN}, 
\newline
QSVC: \url{https://github.com/AakashShindeHelsinki/Data_Reduction_Q-SVM.git},
\newline 
$4\times4$ Image Dataset: \url{https://github.com/AakashShindeHelsinki/4X4ImageDatasetGenerator-.git}

\bibliographystyle{ieeetr}
\bibliography{ref}

\newpage
\appendix
\label{appendix}
\section{Ansatz}

\begin{figure}[!h]
\centering
        \begin{tikzpicture}
        \node[scale=0.8, label = U\_TNN] {
        \begin{quantikz}
        & \gate{R_y(\theta_1)} & \ctrl{1} & \qw \\
        & \gate{R_y(\theta_2)} & \targ{}  & \qw
        \end{quantikz}
        };
        \end{tikzpicture}
\vspace{5pt}
        \begin{tikzpicture}
        \node[scale=0.8, label = U\_13] {
        \begin{quantikz}
        & \gate{R_y(\theta_1)} & \gate{R_z(\theta_3)} & \gate{R_y(\theta_4)} & \ctrl{1} & \qw \\
        & \gate{R_y(\theta_2)} & \ctrl{-1}            & \gate{R_y(\theta_5)} & \gate{R_z(\theta_6)} & \qw
        \end{quantikz}
        };
        \end{tikzpicture}
\vspace{5pt}
        \begin{tikzpicture}
        \node[scale=0.8, label = U\_14] {
        \begin{quantikz}
        & \gate{R_y(\theta_1)} & \gate{R_x(\theta_3)} & \gate{R_y(\theta_4)} & \ctrl{1} & \qw \\
        & \gate{R_y(\theta_2)} & \ctrl{-1}            & \gate{R_y(\theta_5)} & \gate{R_x(\theta_6)} & \qw
        \end{quantikz}
        };
        \end{tikzpicture}
\vspace{5pt}
        \begin{tikzpicture}
        \node[scale=0.7, label = U\_SU4] {
        \begin{quantikz}
        & \gate{U_3(\theta_1, \phi_2, \lambda_3)} & \ctrl{1}  & \gate{R_y(\theta_7)} & \targ{} & \gate{R_y(\theta_9)} & \ctrl{1} & \gate{U_3(\theta_{10}, \phi_{11}, \lambda_{12})} &\qw \\
        & \gate{U_3(\theta_4, \phi_5, \lambda_6)} & \targ{} & \gate{R_z(\theta_8)} & \ctrl{-1} &         & \targ{} & \gate{U_3(\theta_{13}, \phi_{14}, \lambda_{15})} &\qw
        \end{quantikz}
        };
        \end{tikzpicture}
\vspace{5pt}
        \begin{tikzpicture}
        \node[scale=0.7, label = U\_5] {
        \begin{quantikz}
        & \gate{R_x(\theta_1)} & \gate{R_z(\theta_3)} & \gate{R_z(\theta_5)} & \ctrl{1} & \gate{R_x(\theta_7)} & \gate{R_z(\theta_9)} & \qw \\
        & \gate{R_x(\theta_2)} & \gate{R_z(\theta_4)} & \ctrl{-1}            & \gate{R_z(\theta_6)} & \gate{R_x(\theta_8)} & \gate{R_z(\theta_{10})} & \qw
        \end{quantikz}
        };
        \end{tikzpicture}
\vspace{5pt}
        \begin{tikzpicture}
        \node[scale=0.8, label = U\_9] {
        \begin{quantikz}
        & \gate{H} & \ctrl{1} & \gate{R_x(\theta_1)} &\qw \\
        & \gate{H} & \control{} & \gate{R_x(\theta_2)} &\qw
        \end{quantikz}
        };
        \end{tikzpicture}
\vspace{5pt}
        \begin{tikzpicture}
        \node[scale=0.7, label = U\_6] {
        \begin{quantikz}
        & \gate{R_x(\theta_1)} & \gate{R_z(\theta_3)} & \gate{R_x(\theta_5)} & \ctrl{1} & \gate{R_x(\theta_7)} & \gate{R_z(\theta_9)} & \qw \\
        & \gate{R_x(\theta_2)} & \gate{R_z(\theta_4)} & \ctrl{-1}            & \gate{R_x(\theta_6)} & \gate{R_x(\theta_8)} & \gate{R_z(\theta_{10})} & \qw
        \end{quantikz}
        };
        \end{tikzpicture}
\vspace{5pt}        
        \begin{tikzpicture}
        \node[scale=0.8, label = U\_SO4] {
        \begin{quantikz}
        & \gate{R_y(\theta_1)} & \ctrl{1} & \gate{R_y(\theta_3)}& \ctrl{1} & \gate{R_y(\theta_5)}& \qw \\
        & \gate{R_y(\theta_2)} & \targ{}  & \gate{R_y(\theta_4)}& \targ{}  & \gate{R_y(\theta_6)}& \qw
        \end{quantikz}
        };
        \end{tikzpicture}
\vspace{5pt}
        \begin{tikzpicture}
        \node[scale=0.8, label = Pooling] {
        \begin{quantikz}
        & \ctrl{1} & \ctrl{1} & \qw \\
        & \gate{R_z(\theta_1)} & \gate{R_x(\theta_2)} & \qw
        \end{quantikz}
        };
        \end{tikzpicture}
    \caption{Ansatz}
    \label{fig:ansataz}
\end{figure}

\end{document}